\documentclass[dvips]{acta}
\usepackage{supertabular,lscape,epsfig}
\usepackage{amssymb}
\usepackage{amsmath}

\SetPages{0}{0}

\SetVol{58}{2008}

\begin{document}

\begin{Titlepage}
\Title{Planets in Stellar Clusters Extensive Search. V.~Search for
planets and identification of 18 new variable stars in the old open
cluster NGC~188.\footnote{Based on data from
the FLWO 1.2m telescope}}

\Author{M~o~c~h~e~j~s~k~a, B. J.,}{Copernicus Astronomical Center, ul.\
Bartycka 18, 00-716 Warszawa, Poland\\
e-mail: mochejsk@camk.edu.pl}

\Author{S~t~a~n~e~k, K. Z.,}{Department of Astronomy, The Ohio State
University, 140 W.\ 18th Avenue, Columbus, OH 43210, USA\\
e-mail: kstanek@astronomy.ohio-state.edu}

\Author{S~a~s~s~e~l~o~v, D. D., S~z~e~n~t~g~y~o~r~g~y~i, A. H.,
C~o~o~p~e~r, R. L., H~i~c~k~o~x, R. C., H~r~a~d~e~c~k~y, V.}
{Harvard-Smithsonian Center for Astrophysics, 60 Garden St.,
Cambridge, MA~02138, USA\\
e-mail: sasselov, saint, rcooper, rhickox, vhradecky@cfa.harvard.edu}

\Author{M~a~r~r~o~n~e, D. P.}{Jansky Postdoctoral Fellow, National
Radio Astronomy Observatory; Kavli Institute for Cosmological Physics,
University of Chicago, 5640 South Ellis Avenue, Chicago, IL, 60637, USA\\
e-mail: dmarrone@uchicago.edu}

\Author{W~i~n~n, J. N.}{Department of Physics, and Kavli Institute for
Astrophysics and Space Research, Massachusetts Institute of
Technology, Cambridge, MA, USA\\
e-mail: jwinn@cfa.harvard.edu}

\Author{S~c~h~w~a~r~z~e~n~b~e~r~g-C~z~e~r~n~y, A.}
{Astronomical Observatory of Adam Mickiewicz University, ul. Sloneczna
36, 60-286 Poznan, Poland; Copernicus Astronomical Center, ul.\
Bartycka 18, 00-716 Warszawa, Poland\\
e-mail: alex@camk.edu.pl}

\Received{Month Day, Year}
\end{Titlepage}

\Abstract{We have undertaken a long-term project, Planets in Stellar
Clusters Extensive Search (PISCES), to search for transiting planets
in open clusters. In this paper we present the results for NGC~188, an
old, rather populous cluster. We have monitored the cluster for more
than 87 hours, spread over 45 nights. We have not detected any good
transiting planet candidates. We have discovered 18 new variable stars
in the cluster, bringing the total number of identified variables to
46, and present for them high precision light curves, spanning 15
months.}{planetary systems -- binaries: eclipsing -- cataclysmic
variables -- stars: variables: other -- color-magnitude diagrams }

\section{Introduction}
We have undertaken a long-term project, Planets in Stellar Clusters
Extensive Search (PISCES), to search for transiting planets in open
clusters. To date we have published a feasibility study based on one
season of data for NGC~6791 (Mochejska et al.\ 2002, hereafter
Paper~I) and a catalog of 57 variable stars for our second target,
NGC~2158, based on the data from the first observing season (Mochejska
et al.\ 2004, hereafter Paper~II). We have also published the results
of an extensive search for transiting planets in NGC 6791, based on
over 300 hours of observations, spread over 84 nights (Mochejska et
al.\ 2005, hereafter Paper III). We have not detected any promising
candidates, and derived an estimate of 1.5 expected transiting
planets. Our search in NGC 2158 (Mochejska et al.\ 2006, hereafter
Paper IV) produced a candidate transiting low-luminosity object
with eclipse depth of 3.7\% in the $R$ band.

\begin{figure}[htb]
\includegraphics[scale=0.82]{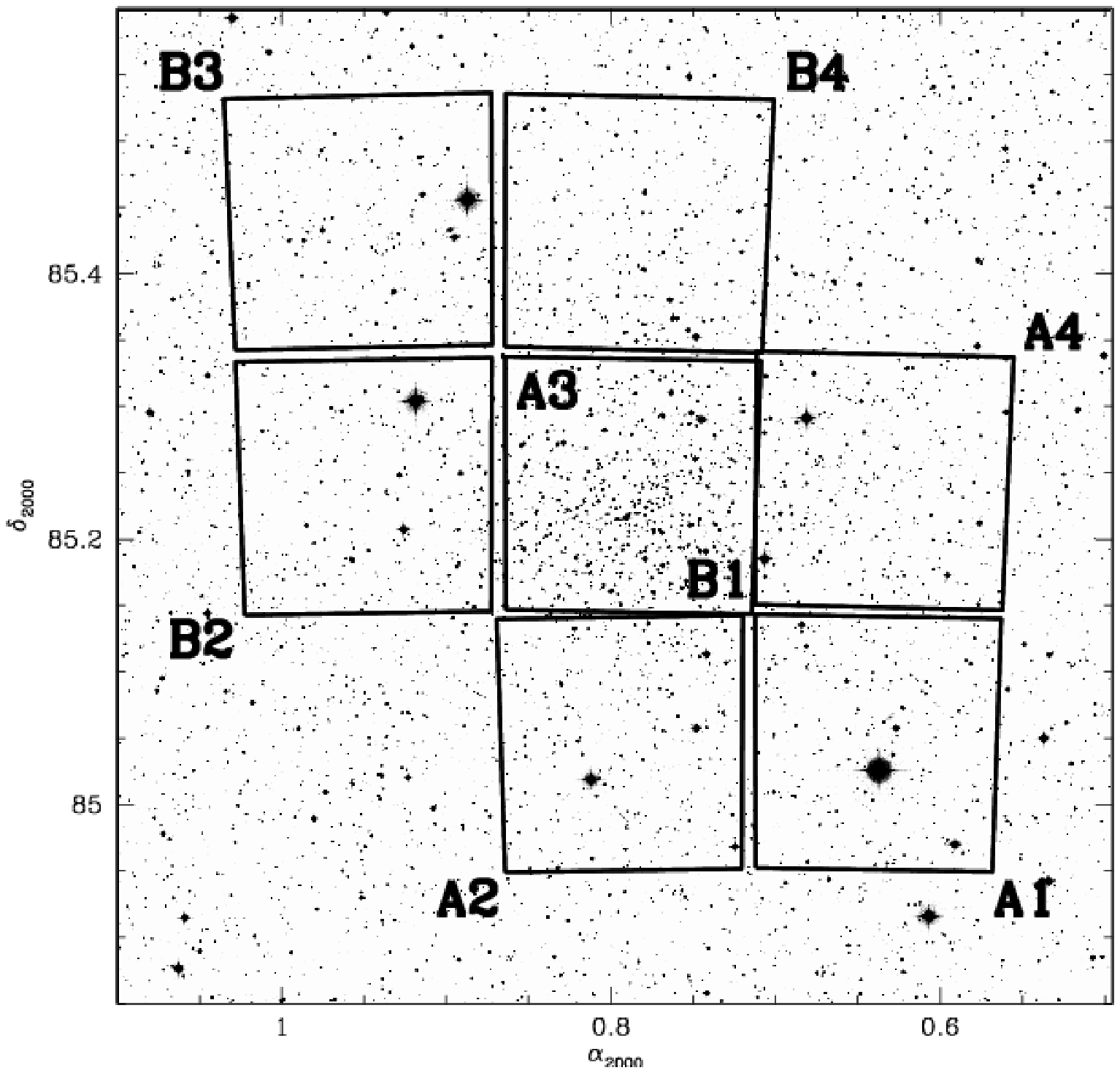}
\FigCap{Digital Sky Survey image of NGC~188 showing fields A and
B. The chips in each field are numbered clockwise from 1 to 4 starting
from the bottom right chip. NGC~188 is centered on chips~A3 and B1.
North is up and east is to the left.}
\end{figure}

In this final paper of the series we present the results of a search
for transiting planets and variable stars in the open cluster NGC~188
$[(\alpha,\delta)_{2000}=(0^h48^m, +85^{\circ}15')$;
$(l,b)=(122^{\circ}.86, +22^{\circ}.38)]$. It is a rather populous,
old ($\tau = 7$ Gyr), solar metallicity ([Fe/H]=-0.04) cluster,
located at a distance modulus of $(m-M)_V=11.44$ (von Hippel \&
Sarajedini 1998; Sarajedini et al.\ 1999).

The paper is arranged as follows: \S 2 describes the observations, \S
3 summarizes the reduction procedure, \S 4 outlines the search
strategy for transiting planets, \S 5 gives an estimate of the
expected number of transiting planet detections and \S 6 contains the
variable star catalog. Concluding remarks are found in \S 7.

\begin{figure}[htb]
\includegraphics[scale=0.67]{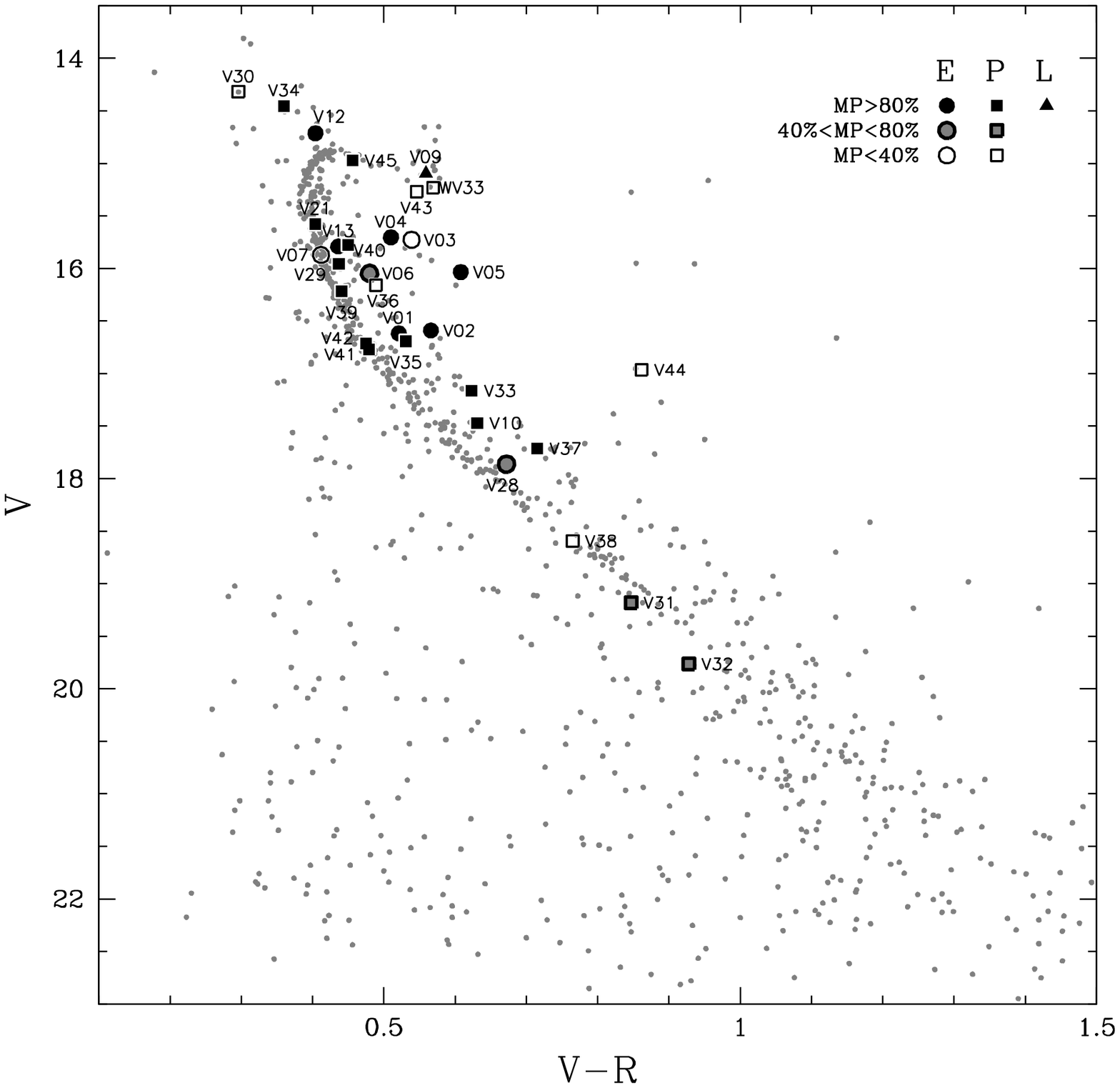}
\FigCap{$V/V-R$ CMD for chip~A3, centered on NGC~188. Eclipsing
binaries are plotted with circles, other periodic variables with
squares and the non-periodic variable with a triangle. Filled
symbols indicate membership probability $MP>80\%$, gray symbols
$40\%<MP<80\%$ and open symbols $MP>40\%$.}
\end{figure}

\begin{figure}[htb]
\includegraphics[scale=0.66]{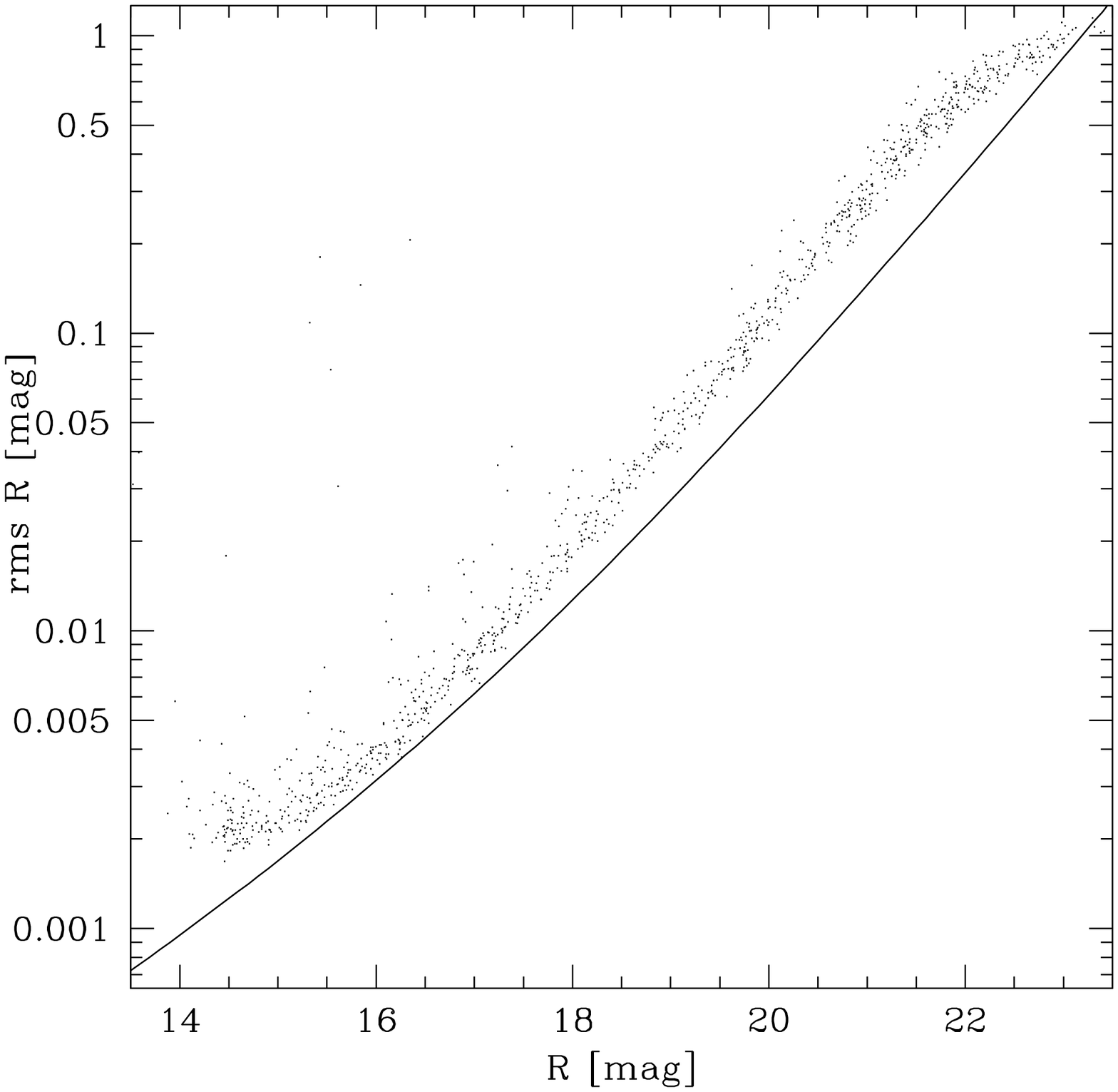}
\FigCap{The rms scatter of the $R$-band light curves for
stars on chip~3, field A, with at least 200 data points. The solid
curve indicates the photometric precision limit due to Poisson noise
of the star and average sky brightness.}
\end{figure}

\section{Observations} 
The data analyzed in this paper were obtained at the Fred Lawrence
Whipple Observatory (FLWO) 1.2 m telescope using the 4Shooter CCD
mosaic with four thinned, back side illuminated AR coated Loral
$2048^2$ CCDs (Szentgyorgyi et al.\ in preparation).  The camera, with
a scale of $0''.33$ pixel$^{-1}$, gives a field of view of
$11'.4\times 11'.4$ for each chip. Two fields, A and B, were
monitored, with the cluster centered on chip~3 in field~A and chip~1
in field~B (Fig.~1). The data were collected during 45 nights, from
2002 September 9 to 2003 December 17. We obtained a total of
$840\times 300$~s exposures in the $R$-band (474 for field A and 366
for B) and $208\times 300$~s exposures in the $V$-band (118 for field
A and 80 for B).

\section{Data Reduction}

\subsection{Image Subtraction Photometry}

The preliminary processing of the CCD frames was performed with the
standard routines in the IRAF ccdproc package.\footnote{IRAF is
distributed by the National Optical Astronomy Observatories, which are
operated by the Association of Universities for Research in Astronomy,
Inc., under cooperative agreement with the NSF.}

Photometry was extracted using the ISIS image subtraction package
(Alard \& Lupton 1998; Alard 2000), as described in detail in Papers~I
and III.

The ISIS reduction procedure consists of the following steps: (1)
transformation of all frames to a common $(x,y)$ coordinate grid; (2)
construction of a reference image from several of the best exposures;
(3) subtraction of each frame from the reference image; (4) selection
of stars to be photometered and (5) extraction of profile photometry
from the subtracted images.

We used the same parameters for image subtraction as in Paper III.
The reference images were constructed from 20 best exposures in $R$
and 15 in $V$.

\subsection{Calibration}

The $VR$ photometry was calibrated against the catalog of Stetson et
al.\ (2004). The zero points in the $R$-band were derived from 44, 99,
547 and 149 stars in chips~A1-A4 and 522, 101, 54 and 133 stars in
chips~B1-B4. The zero points in the $V$-band were derived from 206,
247, 514 and 281 stars in chips~A1-A4 and 495, 251, 222 and 261 stars
in chips~B1-B4. The rms scatter of the residuals was 0.02-0.03 in $R$
on chips with at least 100 stars, and 0.04-0.06 on the remaining three
chips, and 0.02-0.04 in $V$ on all chips.

Figure~2 shows the calibrated V/V-R color-magnitude diagram (CMD) for
the chip~A3 reference image.

\subsection{Astrometry}
Equatorial coordinates were determined for the $R$-band reference
image star lists. The transformation from rectangular to equatorial
coordinates was derived using 149, 182, 518 and 212 transformation
stars in chips~A1-A4, and 506, 190, 161 and 199 stars in chips~B1-B4,
respectively, from the 2MASS catalog (Cutri et al.\ 2003) . The rms
deviation between the catalog and the computed coordinates for the
transformation stars was $1''.0 - 1''.3$ in right ascension (quite
large, due to the closeness of the cluster to the north celestial
pole) and $0''.1$ in declination.

\section{Search for Transiting Planets}

\subsection{Further Data Processing}

We rejected from further analysis 59 field A and 63 field B $R$-band
epochs where the full-width at half-maximum (FWHM) of the stellar
point-spread function (PSF) was greater than 10 pixels and fewer than
1000 stars were detected on chips 3A and 1B by DAOphot (Stetson
1987). We also rejected additional 53 field A and 34 field B bad
quality images.  This left us with 362 and 269 highest quality
$R$-band exposures in fields A and B, with a median seeing of
$2''.2$. We also removed 4 and 5 $V$-band images from fields A and B,
which left us with 114 and 85 exposures with a median seeing of
$2''.6$.

Correcting the light curves for systematic effects using the Tamuz et
al.\ (2005) method alone was insufficient, based on the results of
simulations identical to those described in \S 5.2. We found
significant improvement in the results of the simulations when we also
corrected the data for the presence of offsets between the observing
runs. These offsets are probably caused by the periodic UV flooding of
the CCD camera, which alters its quantum efficiency as a function of
wavelength. Following the procedure used in Paper III, the light
curves were corrected by adding offsets to different runs,
individually for each light curve, so that the median magnitude was
the same during each run. There were four runs, each spanning from 59
to 167 points in field A and 42 to 126 points in field B. Typical
sizes of the offsets on chips A3 and B1 were, respectively, 0.001 and
0.002 for stars below R=18 and 0.008 and 0.009 for stars between R=18
and 19.

Fig.~3 shows the rms scatter of the $R$-band light curves for stars on
chip~A3 with at least 200 data points.  The solid curve indicates the
photometric precision limit due to Poisson noise of the star and
average sky brightness.

\begin{figure}[htb]
\includegraphics[scale=0.32]{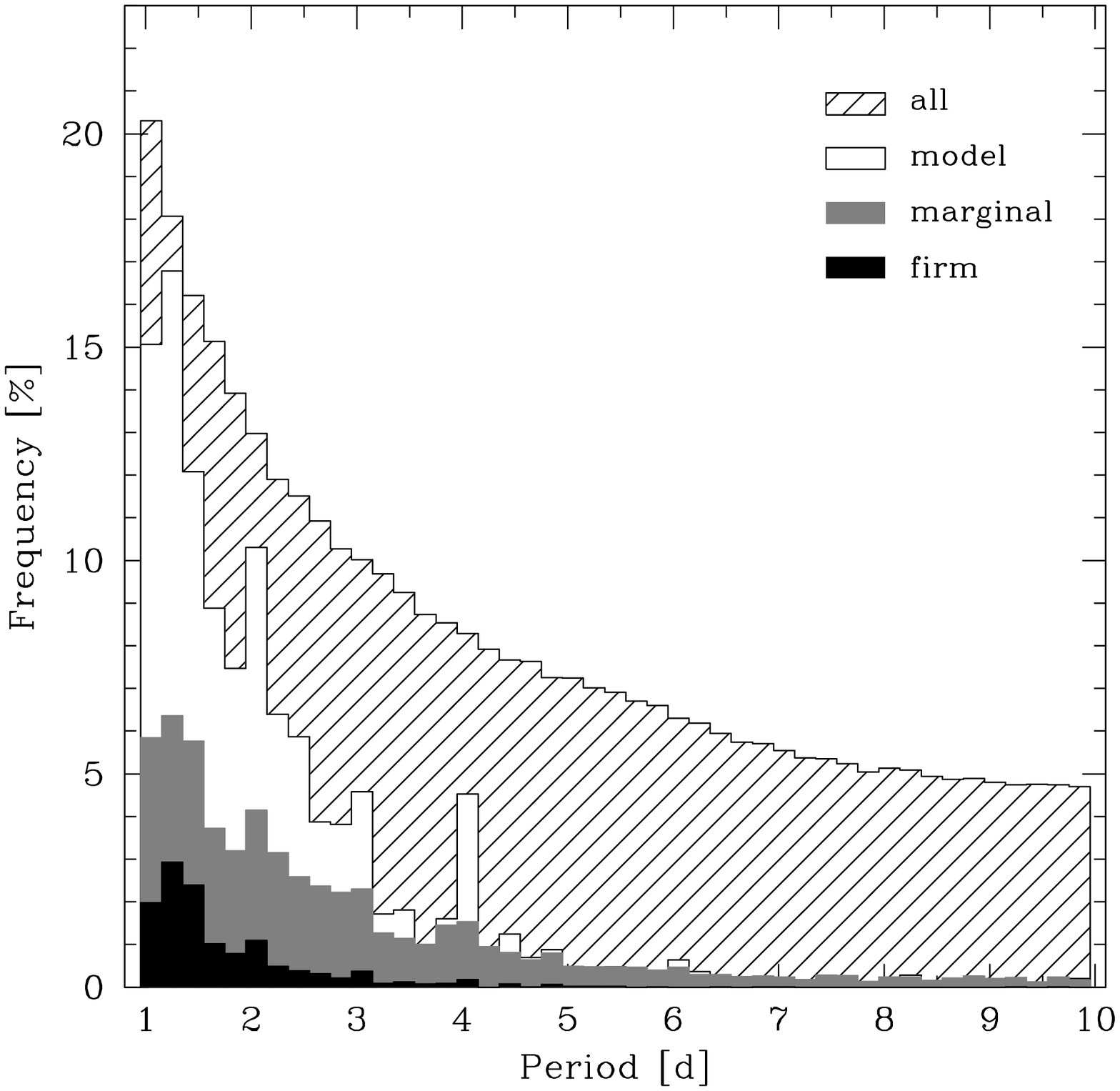}
\includegraphics[scale=0.32]{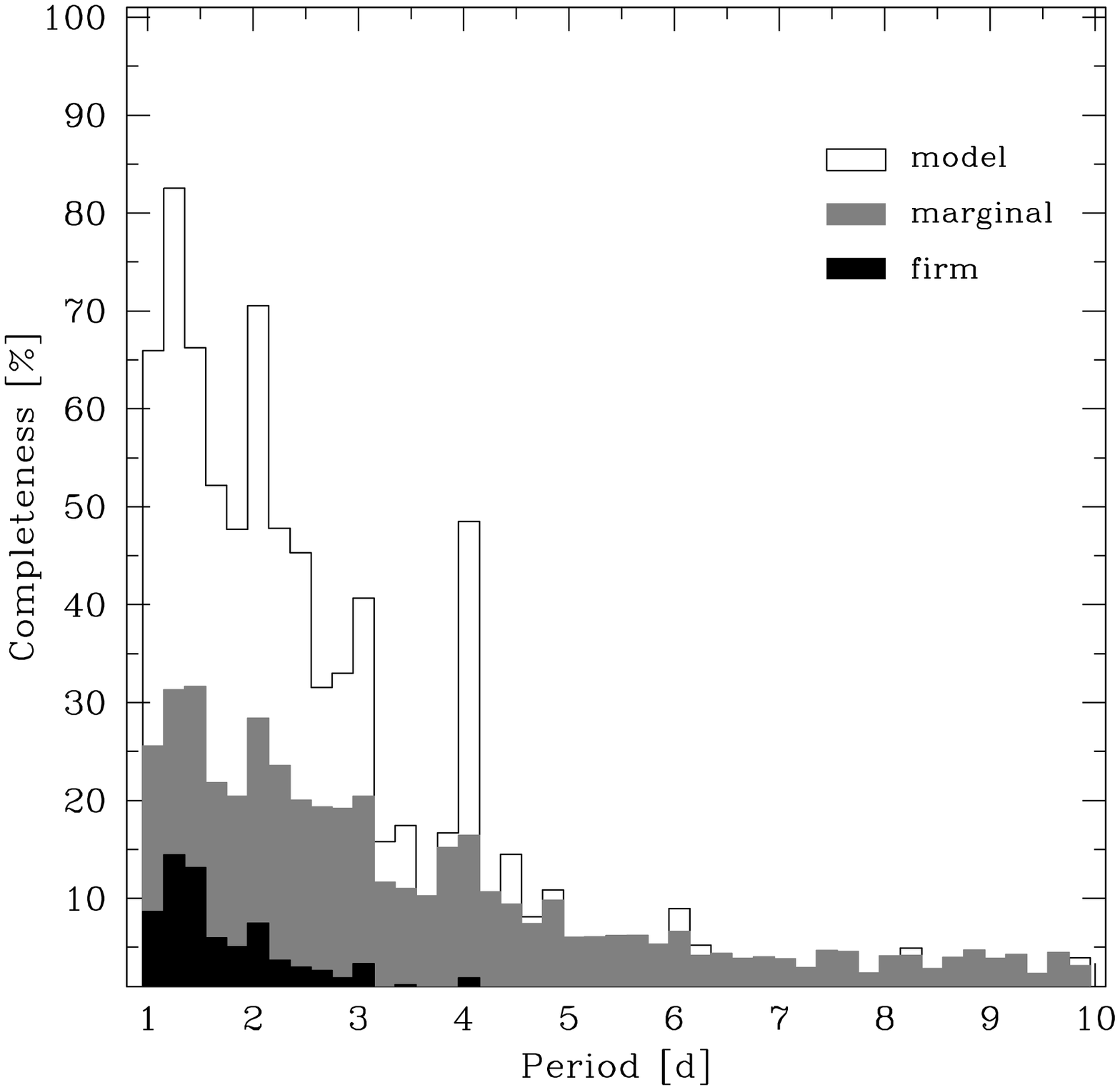}
\FigCap{Detection efficiency of transiting planets as a function of
their period, relative to planets with all inclinations ($left$) and
all transiting planets ($right$). Shown are the distributions for all
transiting planets ({\it hatched histogram}), detections in the model
light curves ({\it open histogram}) and marginal ({\it filled gray
histogram}) and firm ({\it filled black histogram}) detections in the
combined light curves.}
\end{figure}

\subsection{Selection of Transiting Planet Candidates}

For further analysis we selected stars with at least 200 good epochs,
magnitudes $R>14.88$ (the main sequence turnoff; MSTO) and light curve
rms below 0.05 mag. This left us with 2031 stars: 171, 201, 461 and
232 on chips~A1-A4 and 371, 208, 192 and 195 on chips~B1-B4.

To select transiting planet candidates we used the box-fitting
least-squares (BLS) method (Kov{\' a}cs, Zucker, \& Mazeh 2002).
Adopting a cutoff of 5 in Signal Detection Efficiency (SDE) and 9 in
effective signal-to-noise ratio ($\alpha$), we selected 25 stars
on all chips. Among these we did not find any good transiting
planet candidates.

\section{Estimate of the Number of Expected Detections}

To allow for a direct comparison with our previous transiting planet
searches in NGC 6791 (Paper III) and NGC 2158 (Paper IV), we have used
the same method and parameters to estimate the number of expected
detections. We derived the number of transiting planets we should
expect to find, $N_P$, from the following equation:
\begin{equation}
N_P = N_* f_P D
\end{equation}
where $N_*$ is the number of stars with sufficient photometric
precision, $f_P$ is the frequency of planets within the investigated
period range and $D$ is the detection efficiency, which accounts for
random inclinations. ($N_*$). In \S\S 5.1 and 5.2 we determine $f_P$
and $D$.

\begin{figure}[htb]
\includegraphics[scale=0.32]{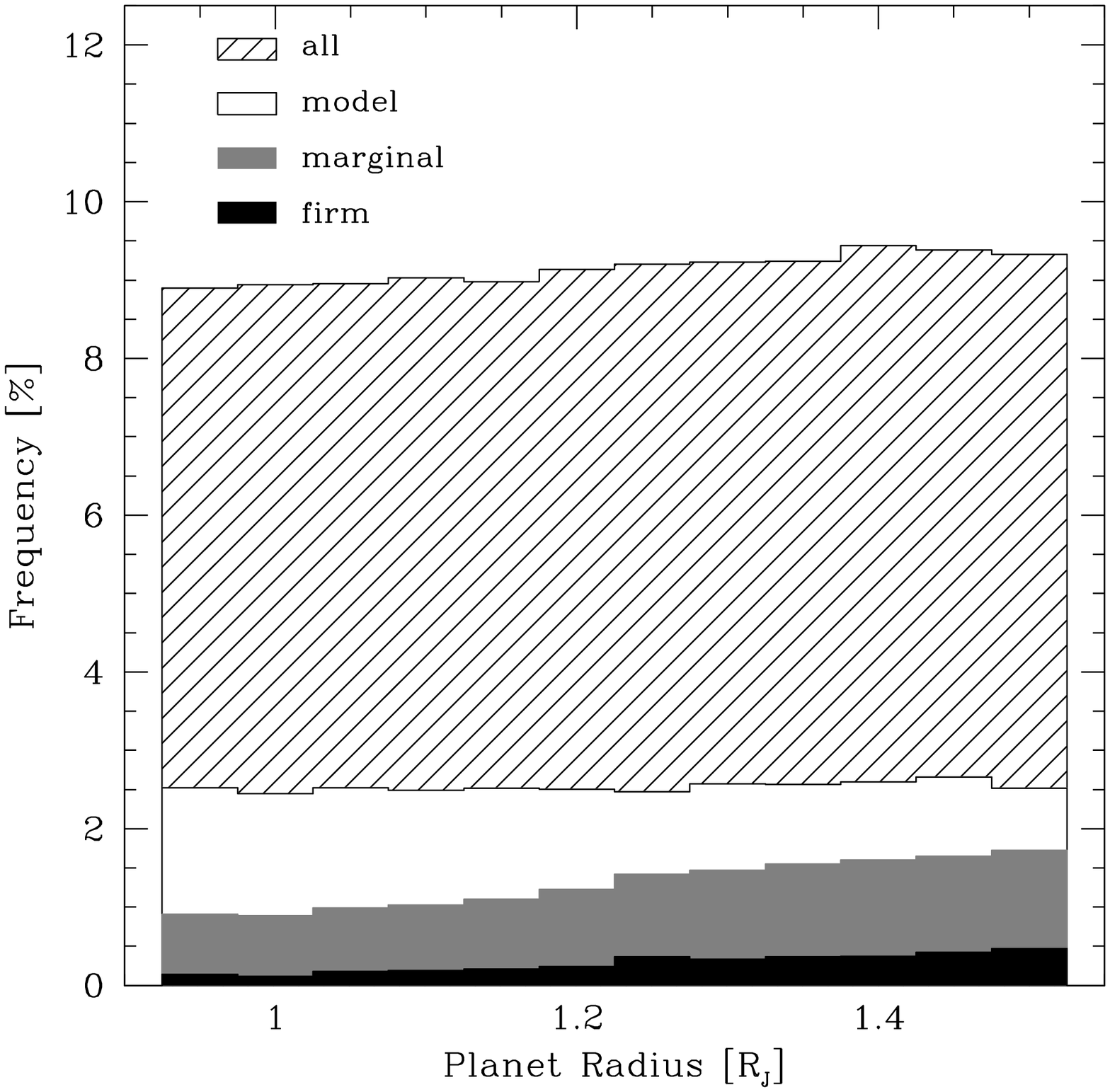}
\includegraphics[scale=0.32]{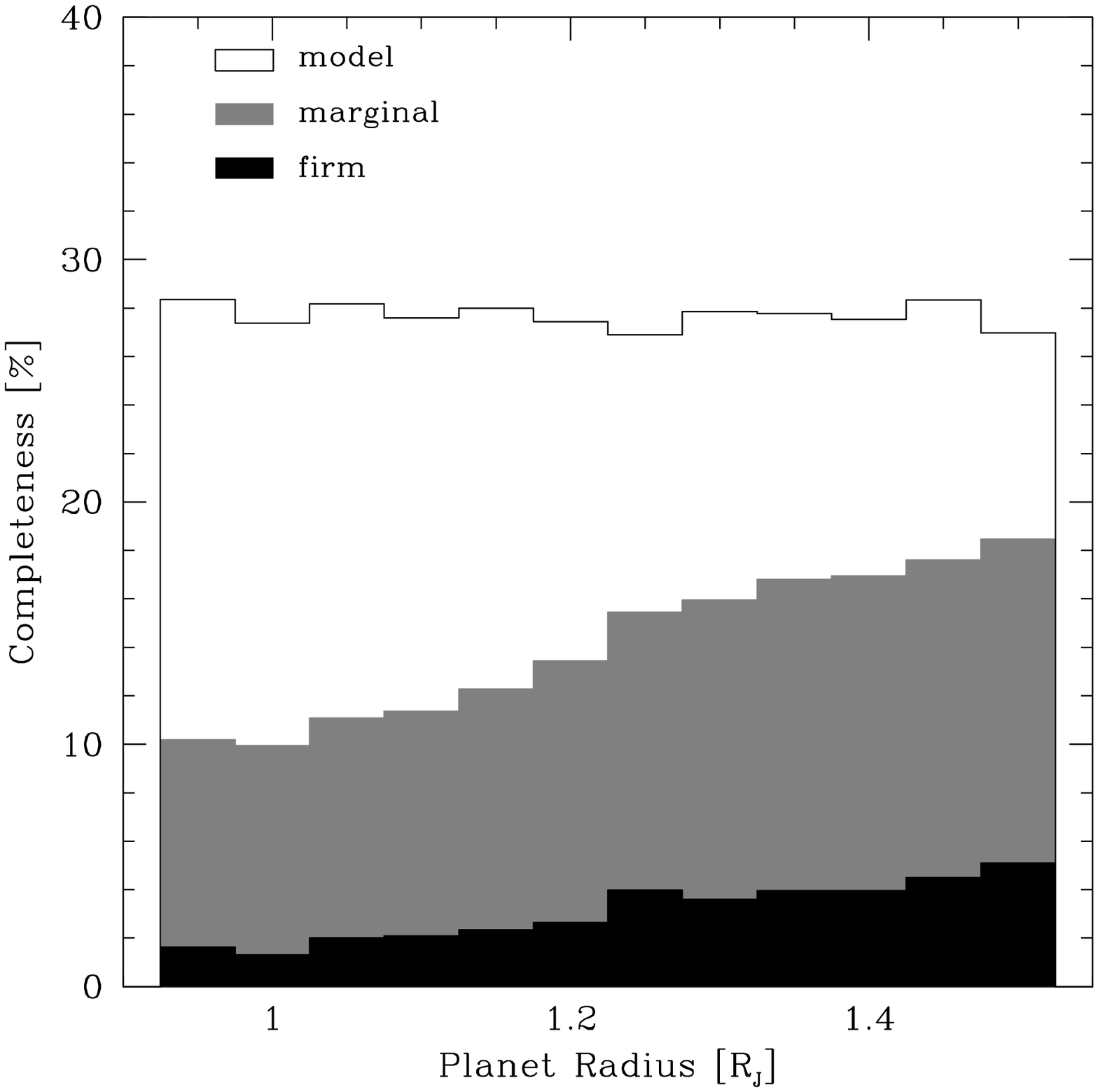}
\FigCap{Detection efficiency of planetary transits as a function of
their radius, relative to planets with all inclinations ($left$)
and all transiting planets ($right$).}
\end{figure}

\subsection{Planet Frequency}

The frequency of planets is known to increase with the host star's
metallicity. From Figure~7 in Santos et al.\ (2004), the frequency of
planets with metallicities below [Fe/H] = +0.1 dex is $\sim$2.5\%.
The percentage of planets with periods below 10 days is $15\%$, as
determined in Paper III. Combining these two numbers yields
$f_P$=0.375\%. 

\subsection{Detection Efficiency}
In order to characterize our detection efficiency, we inserted model
transits into the observed light curves, and tried to recover them
using the BLS algorithm.

\subsubsection{Model Transit Light Curves}

The transit light curves were generated using the approach described
in Paper III. To obtain the radius and mass of the cluster stars, we
used the Z=0.0198, $\tau=7$ Gyr isochrone of Pietrinferni et al.\
(2004). A distance modulus $(m-M)_R=11.34$ mag was used to bring the
observed $R$-band magnitudes to the absolute magnitude scale.

In addition to the period of the planet, $P$, the equations contain
two other free parameters: the planet radius, $R_P$ and the
inclination of the orbit, $i$. A fourth parameter which affects the
detectability of a planet is the epoch of the transits, $T_0$.

\subsection{Test Procedure}

We investigated the range of parameters specified in Table 1, where
$P$ is expressed in days, $R_P$ in $R_J$, $T_0$ as a fraction of
period. We examined the range of periods from 1.05 to 9.85 days and
planet radii from 0.95 to 1.5 $R_J$, with a resolution of $0.2$ days
and $0.05$ $R_J$, respectively. For $T_0$ we used an increment of 5\%
of the period, and a 0.025 increment in $\cos i$. The total number of
combinations is 432000.

We followed the test procedure described in Paper III. The tests were
run on the 2031 stars with at least 200 good epochs and light curve
rms below 0.05 mag.

To assess the impact of the procedure to correct for offsets between
the runs on our detection efficiency, we investigated three cases,
where the correction was applied:
\begin{itemize}
\vspace{-0.2cm}
\item [A.]{after inserting transits,}
\vspace{-0.2cm}
\item [B.]{before inserting transits,}
\vspace{-0.2cm}
\item [C.]{was not applied at all.}
\vspace{-0.2cm}
\end{itemize}
Case (B) will give us the detection efficiency if our data did not
need to be corrected, and case (C) if we did not apply the
corrections. Case (A) will give us our actual detection efficiency,
and its comparison with cases (B) and (C) will show how it is affected
by the applied correction procedure.

\subsection{Detection Criteria}

The same detection criteria as in Paper III were used. A transit was
flagged as detected if:
\begin{enumerate}
\vspace{-0.2cm}
\item {The period recovered by BLS was within 2\% of the input period
$P_{inp}$, 2 $P_{inp}$ or $\frac{1}{2}$ $P_{inp}$,}
\vspace{-0.2cm}
\item {The BLS statistics were above the following thresholds: $SDE >
5$, $\alpha > 9$.}
\vspace{-0.2cm}
\end{enumerate}
These detections will be referred to hereafter as {\it firm}.
Detections where only condition (1) was fulfilled will be called {\it
marginal}.

\subsection{Detection Efficiency}

The results of the tests are summarized in Table~2, which lists the
test type (A-C), the number and percentage of transits with $t_T \geq
0.5^h$ (out of the 432000 possible parameter combinations), and the
numbers and percentages (relative to the total number of transits in
column 2) of transits detected in the model light curves, and of
marginal and firm detections in the combined light curves.

Figures~4 and 5 show the dependence of the detection efficiency on
period and planet radius. The hatched, open, filled gray and filled
black histograms denote distributions for all transiting planets,
planets detected in the model light curves, and marginal and firm
detections in the combined light curves, respectively. Left panels
show the frequency of transits and transit detections relative to
planets with all inclinations. Right panels show the detection
completeness normalized to all transiting planets (plotted as hatched
histograms in left panels).

The tests show that 9\% of planets with periods 1-10 days will transit
their parent stars. This frequency drops from $\sim$20\% at $P=1^d$ to
$\sim$5\% at $P=10^d$. The frequency of transits increases very weakly
with planet radius.

The percentage of detections for the model light curves illustrates
the limitation imposed on our detection efficiency by the temporal
coverage alone. Due to incomplete time sampling, we are restricted to
28\% of all planets with periods between 1 and 10 days. For periods
below 3 days, our temporal coverage is sufficient to detect 57\% of
all transiting planets, and drops to 15\% for periods $3-6$ days and
below 3\% for periods $6-9$ days. The detection completeness does not
depend on the planet radius.

For cases A, B and C, we {\it marginally} detect 14\%, 14\% and 13\%
of all transiting planets, and {\it firmly} detect 3.1\%, 3.3\% and
2.7\%, respectively. Transiting planets with firm detections
constitute 68\%, 69\% and 68\% of all stars with $SDE > 5$ and $\alpha
> 9$. Correcting the light curves after inserting transits (case A)
produces almost the same number of detections, compared to the case
where the correction was applied to the original light curves (case
B). If the light curves were not corrected (case C), we would detect
82\% of the transiting planets detected in case A.

The detection completeness for firm detections peaks at 15\% for
periods $1-1.5^d$ and decreases with period more steeply than model
detections. It strongly increases with increasing planet radius, from
$\sim 1.5\%$ at 1 $R_J$ to 5\% at 1.5 $R_J$.

The efficiency of {\it firm} transiting planet detections, relative to
planets with all orbital inclinations, $D$, is $1232/432000=0.29\%$.

\subsection{Number of Transiting Planets Expected}

In \S\S~5.1-5.2 we determined the planet frequency $f_P$ to be
$0.375\%$, the number of stars as 2031 and our detection efficiency
$D$ as $0.29\%$. We should thus expect 0.02 transiting planets among
the cluster and field stars.

\subsection{Discussion}

Figure~4 demonstrates that our temporal coverage already places
significant limits on our detection efficiency, restricting us to less
than one third of all transiting systems with periods 1-10 days.

After the first observing run for this cluster we realized that this
is not an optimal target for a transiting planet search, due to its
low stellar density. We therefore chose to concentrate on NGC 2158 and
to observe NGC 188 during the time of the night when the former
cluster was unobservable, hence the limited temporal coverage.

\begin{figure}[htb]
\includegraphics[scale=0.71]{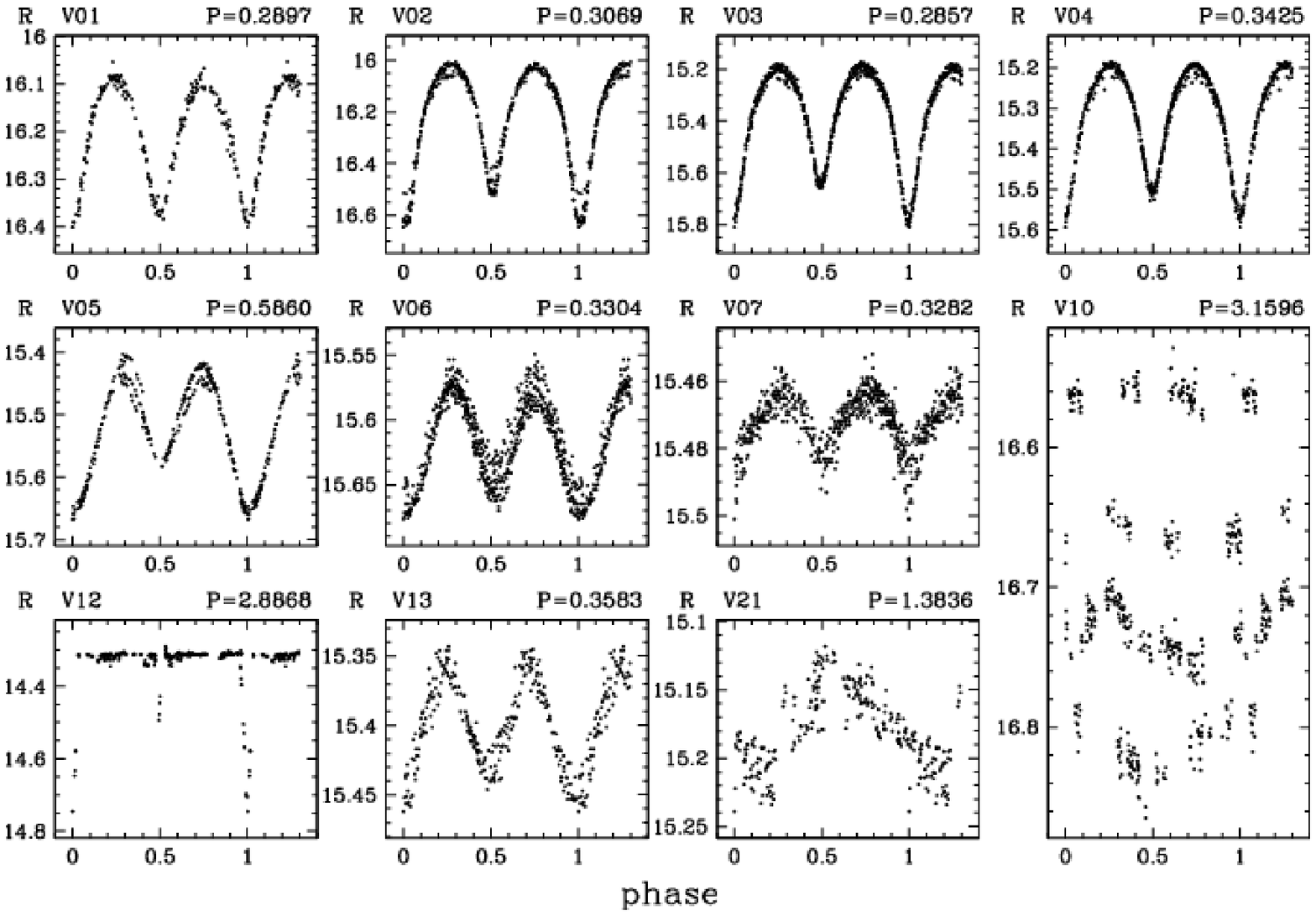}
\FigCap{$R$-band light curves of 11 previously known periodic
variables. Data points from the first observing season are denoted by
gray symbols and from the second season by black symbols. For variable
V10 data from the four observing runs are plotted separately, with
offsets of +0.3, +0.2, +0.1 and 0 mag, respectively.}
\end{figure}

\section{Variable Stars}

Table~3 lists the parameters of the previously known and newly
discovered variable stars in NGC~188. It includes a membership
probability estimate from Platais et al.\ (2003). The light curves of
the variables are shown in Figures~6, 7 and 8.\footnote{The $VR$ band
photometry and finding charts for all variables are available from the
authors from the PISCES website:
http://users.camk.edu.pl/mochejsk/PISCES/.} They are also plotted on
the CMD in Fig.~2.

\begin{figure}[htb]
\includegraphics[scale=0.68]{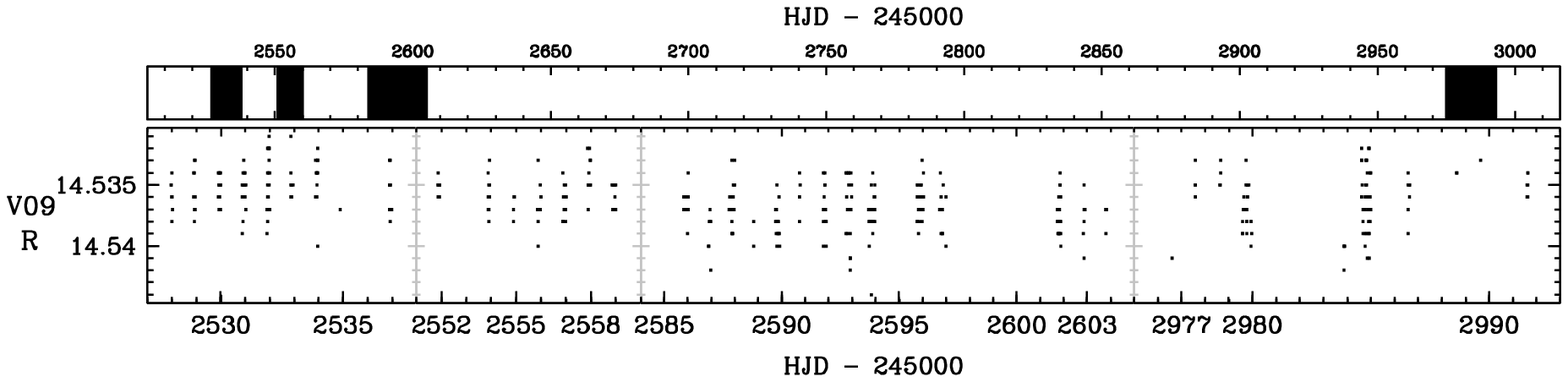}
\FigCap{$R$-band light curve of the previously known non-periodic
variable V9. The top window illustrates the distribution in
time of the four sub-windows plotted for the variable.}
\end{figure}

\begin{figure}[htb]
\includegraphics[scale=0.71]{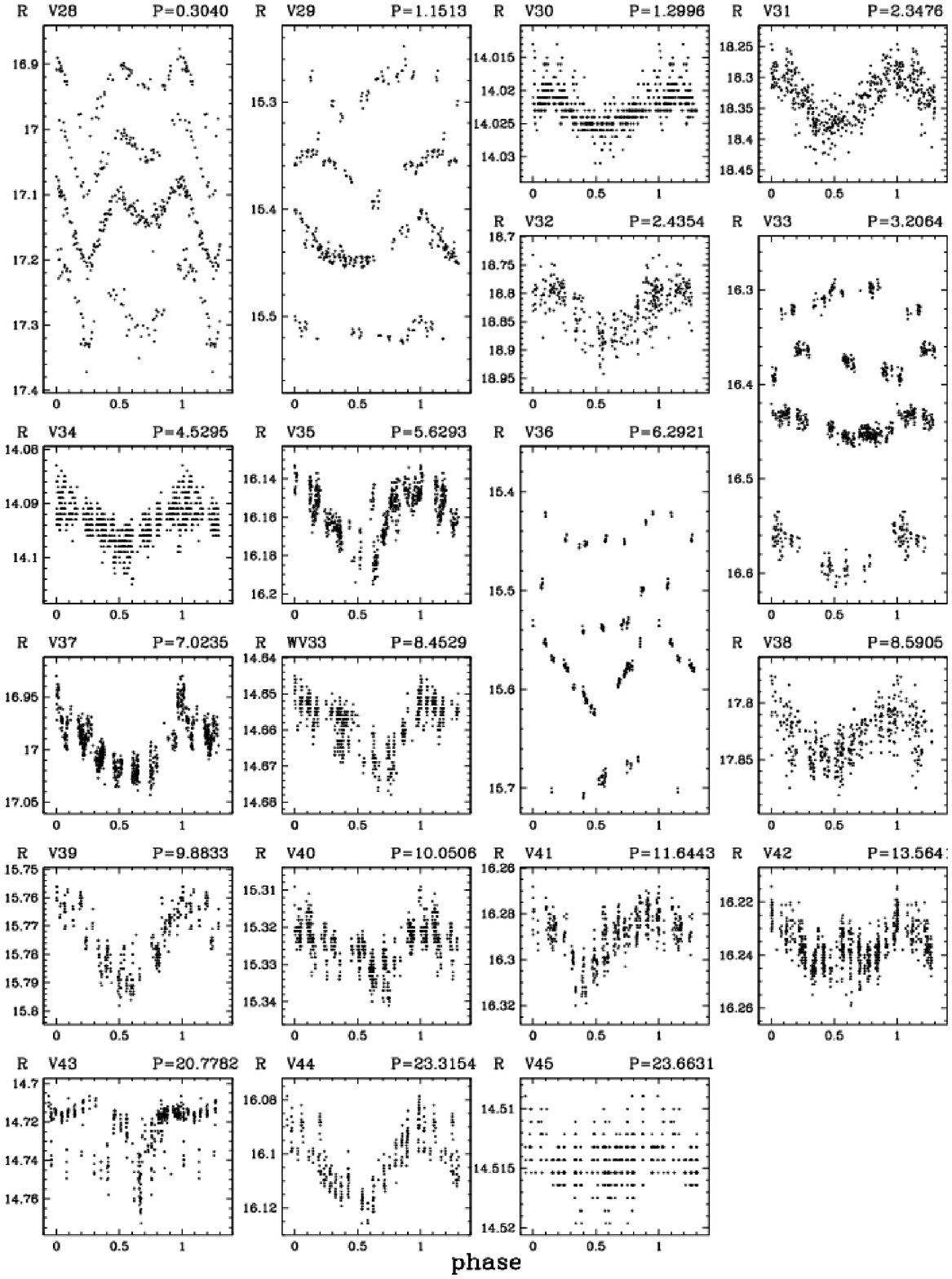}
\FigCap{$R$-band light curves of the previously known variable WV33
and 18 new variables V28-V45. Data points from the first observing
season are denoted by gray symbols and from the second season by black
symbols. For variables V28, V29, V33 and V36 data from the four
observing runs are plotted separately, offset by 0.1, 0.08, 0.08 and
0.08 mag, respectively, between the runs, relative to the last season
data (black points).}
\end{figure}

\subsection{Previously Known Variables}

NGC~188 has been the subject of several variability searches.
Hoffmeister (1964) reported the first four variable stars in this
cluster: EP, EQ, ER and ES Cephei (often referred to as
V1-V4). Kaluzny \& Shara (1987) identified six new variables, V5-V10,
Kaluzny (1990) one, V11, Zhang et al.\ (2002) eight, V12-V19, and
Zhang et al.\ (2004) another eight, bringing the total of known
variables in this cluster to 27.

We identified variables V1-V7, V9, V10, V12, V13, V20 and V21 in our
data. V8 and V11 were saturated on our images. The remaining 12
variables were outside our field of view.

Another study was performed by Kafka \& Honeycutt (2003), who reported
51 variable stars, numbered WV1 through WV51. They identified their
variables WV1, WV14 and WV11 with the previously known variables V1,
V2 and V3. We identified 32 of their newly discovered variables in our
fields: WV2-WV10, WV12, WV13, WV16-WV18, WV20-WV25, WV29, WV32-WV35,
WV38, WV39, WV44-WV46, WV48, WV50. Two other variables, WV40 and WV41,
were within our fields of view but we did not have photometry for
them. We have examined their light curves and 31 of them did not
display any kind of variability in our data. One variable, WV33, was
found to vary with a period 5.5 times longer than the one reported by
the authors. We also note that the coordinates for variables WV27,
WV41 and WV44 disagree with their positions on the finder charts and
that variable WV47 does not have any coordinates listed and it is not
present on the finder charts.

Remarks on particular variables:

V1-V7: These variables are W UMa type contact binaries. We confirm
their periods reported in the literature. We observe some variability
in the shape of the light curves, especially for V5 and V6. V1, V2, V4
and V5 have high cluster membership probabilities ($MP>82\%$), V6 may
be a cluster member ($MP=71\%$) and V3 and V7 are probably non-members
($MP=0\%$ and 12\%, respectively). All of these stars, regardless of
cluster membership, appear to be located on the cluster main sequence
(MS) or above it, on the binary sequence.

V9: This variable shows very weak variability in our data. It is a
cluster member ($MP=97\%$), located on the subgiant branch.

V10: This is a cluster member with probability 94\%, located on the
binary sequence. The phase, amplitude and the magnitude zero point of
V10 are seen to vary between the observing runs. Kaluzny (1990) and
Zhang et al.\ (2004) noted that in their data this star exhibited weak
or no variability. This confirms that the variability amplitude of
this star does indeed change with time. Kafka \& Honeycutt (2003)
determined a period of 3.1 days for this star, based on data from
Kaluzny \& Shara (1987), and we have found a similar period of 3.15
days in our observations. This is most likely a BY Dra type variable,
where the variability is due to large spots on the star's surface
rotating in and out of view.

V12: This is a detached eclipsing binary in the cluster ($MP=97\%$),
located above the MSTO -- a blue straggler. We have observed
two eclipses and derive a period of 2.89 days, similar to the
period of 2.81 days suggested by Zhang et al.\ (2004).

V13: This is a W UMa contact binary, a cluster member ($MP=98\%$),
located above the MS. We derive a slightly longer period than
Zhang et al.\ (2002; 2004). 

V20: This star does not display variability in our data.

V21: This is a cluster member ($MP=92\%$), located on the MS. Our data
do not confirm the period of 1.17 days found by Zhang et al.\
(2004). We find a period of 1.38 days, assuming that this is a BY Dra
type variable. A period of 2.77 days would be required to make this a
W UMa system, as suggested by Zhang et al.\ (2004) -- a rather
unlikely scenario. A change in the variability amplitude is seen for
this star.

WV33: This star is most likely not associated with the cluster
($MP=0\%$). It seems to be a BY Dra type variable with a period of 8.5
days. Our data exclude the period of 1.54 days suggested by Kafka \&
Honeycutt (2003).

\subsection{New Variables}

We searched for new variables by running BLS and Tatry
(Schwarzenberg-Czerny \& Beaulieu 2006) on the light curves of 6845
stars in fields A and B in the period ranges of 0.1-10 days and
0.1-100 days, respectively. We have discovered 18 new variables: 1, 3,
10 and 1 in field A chips~1-4 and 1 and 2 in field B chips 2 and 3. We
have recovered 8 of the new 10 variables from chip A3 on chip B1. The
remaining two variables were outside of the field of view. Of these
new variables one is an eclipsing binary and 17 are other periodic
variables.

V28 is a contact eclipsing binary and it displays long-term changes in
the shape of the light curve. On the CMD it is located just above the
MS. It may be a member of the cluster ($MP=59\%$).

The remaining 17 periodic variables display roughly sinusoidal light
variations with periods ranging from 1.3 to 24 days. Most of them
display BY Dra type of variability. V29, V33, V36 and V43, in addition
to periodic variability, show a change in the variation amplitude,
phase and magnitude zeropoint between the observing runs, which is
most likely caused by the evolution of the spots on their surface.

Variables V29, V33-V35, V37, V39-V42 and V45 are likely cluster
members, with $MP$ ranging from 91\% to 98\%. V29, V39, V41 and V42
are located on the cluster MS. V33, V35, V37, V40 are located above
the MS, on the cluster binary sequence, V45 is located on the subgiant
branch and V34 is a blue straggler. For V31 and V32, the faintest
variables in our sample, the membership probability estimates are
inconclusive (53\% and 46\%, respectively).  They are located near the
cluster MS. V30, V36, V38, V43 and V44 are most likely not members of
the cluster ($MP=0\%$).

\section{Conclusions}
In this paper we have performed a search for transiting planets in the
old cluster NGC~188. The cluster was monitored for over 87 hours
during 45 nights. Assuming a planet frequency from radial velocity
surveys, we estimate that we should have detected 0.02 transiting
planets with periods between 1 and 10 days, with our photometric
precision and temporal coverage.

We have discovered 18 new variable stars in the old open cluster NGC
188: one eclipsing binary and 17 other periodic variables. Together
with 28 previously identified variables, this brings the total number
of known variables in this cluster to 46. We have also presented high
photometric precision light curves, spanning almost 15 months, for all
previously known variables within our field of view.

\Acknow{ We would like to thank the FLWO 1.2 m Time
Allocation Committee for the generous amount of time we were allocated
to this project. 

Support for BJM was provided by the Polish KBN grant P03D012.30.

This publication makes use of data products from: the Two Micron All
Sky Survey, which is a joint project of the University of
Massachusetts and the Infrared Processing and Analysis
Center/California Institute of Technology, funded by the National
Aeronautics and Space Administration and the National Science
Foundation; the Digital Sky Survey, produced at the Space Telescope
Science Institute under U.S. Government grant NAG W-2166; the SIMBAD
database, operated at CDS, Strasbourg, France and the WEBDA open
cluster database maintained by J.~C.~Mermilliod
(http://obswww.unige.ch/webda/).}

\MakeTable{lllcc}{12.5cm}{Parameter Range}
{\hline
Parameter & min & max & step & n$_{steps}$\\
\hline
P (days)      &  1.05   &  9.85 &  0.200 &  45\\
$R_P$ ($R_J$) &  0.95   &  1.50 &  0.050 &  12\\
$T_0$         &  0.00   &  0.95 &  0.050 &  20\\
$\cos i$      &  0.0125 &  0.9875 &  0.025 &  40\\
\hline
}

\MakeTable{crrrrrrrr}{12.5cm}{Artificial transit test statistics}
{\hline
test & \multicolumn{2}{c}{all transits} &
\multicolumn{2}{c}{model} &
\multicolumn{2}{c}{marginal} &
\multicolumn{2}{c}{firm}\\
type & N & \% & N & \% & N & \% & N & \% \\
\hline
 1 & 39511 &    9.1 & 10940 &   27.7 &  5603 &   14.2 &  1232 &    3.1\\
 2 & 39513 &    9.1 & 10939 &   27.7 &  5723 &   14.5 &  1300 &    3.3\\
 3 & 39511 &    9.1 & 10940 &   27.7 &  5124 &   13.0 &  1060 &    2.7\\
\hline
}

\MakeTable{rrrrrrrrcc}{12.5cm}{Variable stars in NGC 188}
{\hline
ID & $\alpha_{2000}$ [h] & $\delta_{2000}$ [$\circ$] & P [d] &
$R$ & $V$ &$A_R$ & $A_V$ & $Type$ & $MP$\\
\hline
 V03 & 00 50 27.69 & 85 15 08.42 &  0.2857 & 15.188 & 15.727 &  0.608 &  0.557 & E &  0 \\
 V01 & 00 46 54.02 & 85 21 43.58 &  0.2897 & 16.096 & 16.617 &  0.290 &  0.276 & E & 98 \\
 V28 & 00 52 08.82 & 85 19 05.97 &  0.3040 & 17.190 & 17.862 &  0.145 &  0.126 & E & 59 \\
 V02 & 00 47 33.37 & 85 16 24.14 &  0.3069 & 16.025 & 16.591 &  0.604 &  0.562 & E & 82 \\
 V07 & 00 46 14.30 & 85 13 59.28 &  0.3282 & 15.460 & 15.872 &  0.036 &  0.043 & E & 12 \\
 V06 & 00 47 16.44 & 85 15 35.31 &  0.3304 & 15.565 & 16.045 &  0.109 &  0.095 & E & 71 \\
 V04 & 00 50 50.01 & 85 16 11.66 &  0.3425 & 15.193 & 15.703 &  0.378 &  0.360 & E & 96 \\
 V13 & 00 51 14.99 & 85 24 51.17 &  0.3583 & 15.355 & 15.791 &  0.100 &  0.088 & E & 98 \\
 V05 & 00 48 22.53 & 85 15 54.72 &  0.5860 & 15.426 & 16.034 &  0.232 &  0.220 & E & 95 \\
 V29 & 00 52 45.98 & 85 12 14.47 &  1.1513 & 15.519 & 15.956 &  0.026 &  0.026 & P & 98 \\
 V30 & 00 46 53.98 & 85 14 37.03 &  1.2996 & 14.023 & 14.319 &  0.005 &  0.003 & P &  0 \\
 V21 & 00 50 02.82 & 85 21 22.91 &  1.3836 & 15.173 & 15.577 &  0.067 &  0.069 & P & 92 \\
 V31 & 00 49 25.38 & 85 01 18.10 &  2.3476 & 18.335 & 19.182 &  0.090 &  0.119 & P & 53 \\
 V32 & 00 54 20.04 & 85 24 01.20 &  2.4354 & 18.836 & 19.764 &  0.090 &  0.124 & P & 46 \\
 V12 & 00 52 37.71 & 85 10 34.61 &  2.8868 & 14.308 & 14.712 &  0.334 &  0.281 & E & 97 \\
 V10 & 00 50 44.62 & 85 11 38.93 &  3.1596 & 16.840 & 17.471 &  0.033 &  0.028 & P & 94 \\
 V33 & 00 47 04.29 & 85 15 01.54 &  3.2064 & 16.540 & 17.163 &  0.025 &  0.027 & P & 93 \\
 V34 & 00 47 11.82 & 85 13 31.46 &  4.5295 & 14.094 & 14.454 &  0.007 &  0.007 & P & 98 \\
 V35 & 00 45 18.64 & 85 18 36.90 &  5.6293 & 16.161 & 16.692 &  0.039 &  0.040 & P & 97 \\
 V36 & 00 55 43.58 & 85 24 00.85 &  6.2921 & 15.669 & 16.158 &  0.055 &  0.061 & P &  0 \\
 V37 & 00 44 14.11 & 85 16 33.89 &  7.0235 & 16.999 & 17.714 &  0.057 &  0.052 & P & 91 \\
WV33 & 00 51 15.85 & 85 09 47.86 &  8.4529 & 14.660 & 15.229 &  0.018 &  0.021 & P &  0 \\
 V38 & 00 44 30.63 & 85 01 29.68 &  8.5905 & 17.830 & 18.595 &  0.038 &  0.090 & P &  0 \\
 V39 & 00 39 05.98 & 85 18 39.32 &  9.8833 & 15.776 & 16.217 &  0.027 &  0.025 & P & 91 \\
 V40 & 00 44 52.20 & 85 15 54.34 & 10.0506 & 15.326 & 15.776 &  0.013 &  0.016 & P & 98 \\
 V41 & 00 49 36.46 & 85 06 26.08 & 11.6443 & 16.290 & 16.769 &  0.025 &  0.031 & P & 96 \\
 V42 & 00 48 54.98 & 85 17 12.61 & 13.5641 & 16.237 & 16.712 &  0.017 &  0.016 & P & 91 \\
 V43 & 00 36 39.97 & 85 03 18.20 & 20.7782 & 14.724 & 15.270 &  0.025 &  0.012 & P &  0 \\
 V44 & 00 48 25.86 & 85 12 23.35 & 23.3154 & 16.103 & 16.965 &  0.026 &  0.041 & P &  0 \\
 V45 & 00 44 40.60 & 85 15 21.22 & 23.6631 & 14.514 & 14.970 &  0.002 &  0.004 & P & 98 \\
 V09 & 00 44 52.25 & 85 10 25.01 &  ... & 14.536 & 15.095 &  0.011 &  0.016 & L & 97 \\
\hline
\multicolumn{10}{p{9cm}}{Cluster membership probabilities $MP$ taken from Platais et al. (2003).}
}

\end{document}